# Quasiparticle Interference in LiFeAs: Signature of Inelastic Tunneling through Spin Fluctuations

Shun Chi,[1,2,*] Carolina A. Marques,[3] W. N. Hardy,[1,2] Ruixing Liang,[1,2] Pinder Dosanjh,[1,2] D. A. Bonn,[1,2] S. A. Burke,[1,2,4] and Peter Wahl[3,5,†]

[1]*Department of Physics and Astronomy,*
*University of British Columbia, Vancouver BC, Canada V6T 1Z1*
[2]*Stewart Blusson Quantum Matter Institute,*
*University of British Columbia, Vancouver BC, Canada V6T 1Z4*
[3]*SUPA, School of Physics and Astronomy,*
*University of St Andrews, North Haugh,*
*St Andrews, Fife, KY16 9SS, United Kingdom*
[4]*Department of Chemistry, University of British*
*Columbia, Vancouver BC, Canada V6T 1Z1*
[5]*Physikalisches Institut, Universität Bonn,*
*Nussallee 12, 53115 Bonn, Germany*
(Dated: September 11, 2025)

## Abstract

Quasi-particle interference (QPI) is a powerful tool to characterize the symmetry of the superconducting order parameter in unconventional superconductors, by mapping the spatial dependence of elastic tunneling of electrons between the tip of a scanning tunneling microscope and a sample. Here, we consider the influence of inelastic tunneling on quasi-particle interference, exemplarily for the iron-based superconductor LiFeAs. We clearly observe replica features in both experimental QPI maps and the dispersion extracted from QPI, which from comparison with theoretical model calculations can be attributed to inelastic tunneling. Analysis of the QPI dispersion shows that the inelastic mode that gives rise to these replica features exhibits a resonance between 8 and 10 meV. Comparison of the energy scale of the resonance energy estimated from QPI with inelastic neutron scattering indicates that the replica features arise from interaction with spin fluctuations.

# INTRODUCTION

Since its invention, Scanning tunneling microscopy (STM) has become an important technique for probing quantum materials and unconventional superconductivity. The local density of states (LDOS) of the sample can be extracted from the elastic tunneling (ET) conductance for both filled and empty states with picometer spatial precision and sub-meV energy resolution. This has provided important insights into the physics of unconventional superconductors, including the iron-based [1] and cuprate [2, 3] superconductors, as well as heavy fermion materials[4, 5]. Complementary to many other spectroscopic techniques, STM is now widely used for measuring the spatial variation of the LDOS both below and above the Fermi energy ($E_\text{F}$) [1, 6]. In certain circumstances, inelastic tunneling processes also contribute to the overall conductance signal due to the interaction between the tunneling electrons and bosonic modes. This inelastic tunneling (IET) contribution has been used to characterize both molecular vibrations [7], spin-flip excitations[8], and spin-spin interactions [9]. Theoretical models and experiments suggest that inelastic excitation of spin fluctuations in strongly correlated electron materials also contribute to the tunneling current [10–12], and may even be a non-negligible contribution.

The contribution from inelastic tunneling can significantly alter the proportionality between the tunneling spectrum $g(V)$ and the LDOS [13–15]. Taking superconducting LiFeAs as an example, Fig. 1(a) shows theoretical simulations of the tunneling spectrum with and without inelastic tunneling via spin excitations, calculating the conductance as originally proposed by Kirtley and Scalapino[10]. There are several notable differences between spectra with and without inelastic contribution included: (1) the tunneling spectrum with IET, $g_\text{tot}(V)$, is V-shaped with its minimum near $E_\text{F}$. This type of V-shape spectrum has been seen in both cuprates and iron-based superconductors [16–18]. (2), as indicated by the red arrows, two kinks appear in the spectrum. When the superconducting spectrum is normalized by the spectrum above the superconducting transition, the kinks become dip-hump structures that scale with the energy of the bosonic mode. This has been observed in many high-$T_c$ superconductors [16, 18, 19]. (3), as indicated by the blue arrow, the top of a hole band results in a step in the LDOS and the elastic tunneling spectrum but appears as a dip in the tunneling spectrum when the inelastic channel is included. As shown in Fig. 1(b), the experimental tunneling conductance of LiFeAs is V-shaped and has two *kinks*. In addition, the plateau from 35 mV to 45 mV is consistent with the upper band edge of hole bands as extrapolated from ARPES [20–22]. It appears as a plateau instead of a dip in Fig. 1(b) due to broadening.

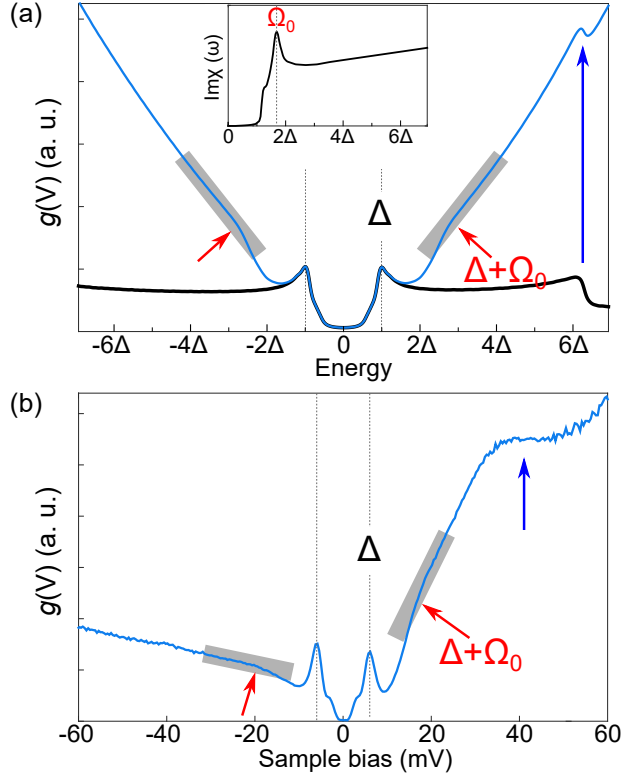

FIG. 1. **Elastic and inelastic contributions to the tunneling spectrum.** (a) Simulated tunneling conductance $g_{el}(V)$ for LiFeAs due to elastic tunneling processes (black) and total tunneling conductance $g_{tot}(V)$ (blue) with the inelastic contribution included, using a five-orbital model and a spin-fluctuation spectrum with a resonance peak at $\Omega_0$ (inset). (b) Experimental tunneling conductance measured at $T = 1.5$ K. The red arrows in (a) and (b) indicate the kinks at $\sim \Delta + \Omega$. The blue arrow highlights the structure in $g_{tot}(V)$ due to the upper band edge of a hole band in the LDOS (step in the green curve).

Even though these dramatic effects in tunneling spectra have been well characterized, the influence of inelastic tunneling on quasiparticle interference (QPI) has rarely been addressed in detail. In this report, we show that, in LiFeAs, inelastic tunneling gives rise to replica features not only in the tunneling spectra, but also in the dispersion observed in QPI. The replica due to IET disperse similar to the signal from the elastic channel but are shifted up (down) in energy by the bosonic resonance energy in the positive (negative) bias side.

Our observations open a new avenue for investigating the interaction between electronic quasiparticles and bosonic excitations.

**RESULTS**

**Theoretical modelling of inelastic contribution to QPI**

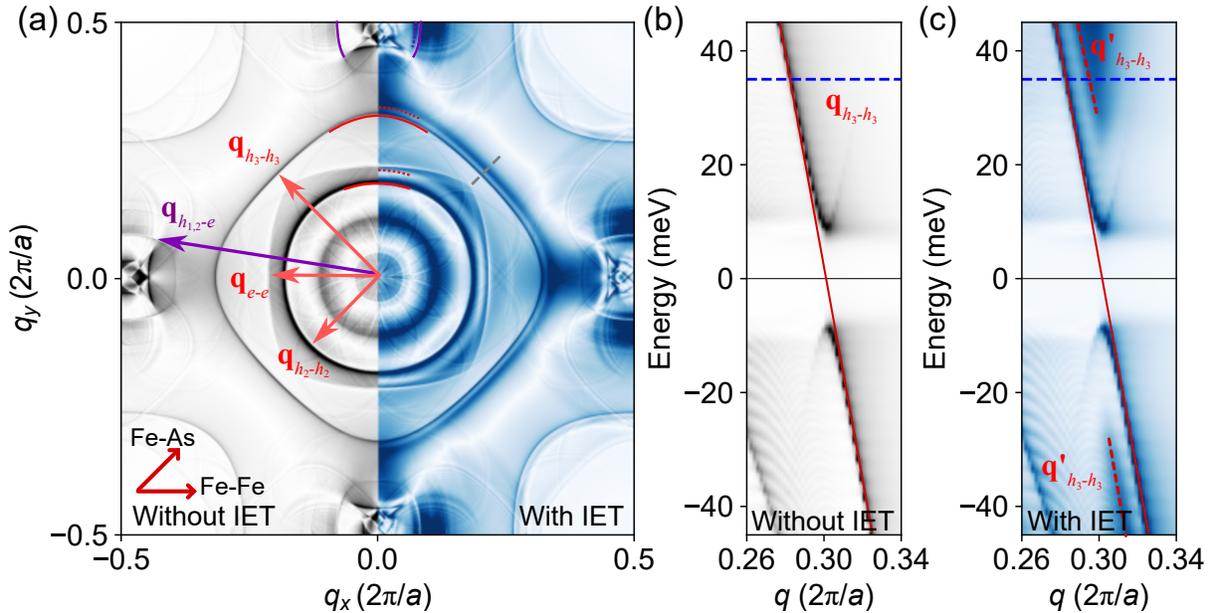

FIG. 2. **Model calculations of inelastic contribution to QPI.** Simulated QPI: (a) $g_{el}(\mathbf{q}, eV)$ (left half) and $g_{tot}(\mathbf{q}, eV)$ (right half) at $eV \sim 3\Delta$ (35 meV). Red arrows highlight intra-band scattering vectors, $\mathbf{q}_{h_{1,2}-h_{1,2}}$ (mix of $\mathbf{q}_{h_1-h_1}$ and $\mathbf{q}_{h_2-h_2}$), $\mathbf{q}_{h_3-h_3}$, and $\mathbf{q}_{e-e}$. The purple arrow indicates the inter-band scattering, $\mathbf{q}_{h_{1,2}-e}$. Red and purple curves indicate elastic QPI features and their inelastic replicas. (b) Cut through $g_{el}(\mathbf{q}, eV)$ along $(0,0) - (\pi, \pi)$ (Fe-As direction, gray dashed line in (a)), showing the dispersion of $\mathbf{q}_{h_3-h_3}$. (c) Same cut as (b) but through $g_{tot}(\mathbf{q}, eV)$, in addition to the dispersion from the normal QPI signal, a replica feature can be seen.

In the IET process, instead of tunneling between states with the same energy between the sample and the tip, quasi-particles will interact with a bosonic excitation, either emitting or absorbing an excitation with energy $\Omega$ during the tunneling process. At very low temperatures, when none of the bosonic excitations are thermally excited, typically the electrons will emit a bosonic excitation and therefore lose energy in the process, tunneling to states that are closer to $E_\text{F}$. For a discrete excitation, this would create replica of the $\omega(q)$ structure shifted by $\Omega$. In the case of a continuous excitation with a sharp onset, states at all energies above the onset $\Omega$ contribute, hence, QPI features associated with $\omega - \Omega$ and below appear at $\omega$ provided that $|\omega| > |\Omega|$, i.e. the energy of the tunneling electrons $\omega$ is larger than that of the excitations $\Omega$. At low energies, where the spectral weight of the bosonic excitations is minimal, only QPI from the elastic channel, $g_\text{el}(\mathbf{q}, eV)$, is present. If the bosonic excitation spectrum exhibits a resonance at an energy $\Omega_0$, when the energy approaches $\Delta + \Omega_0$, the IET channel makes a significant contribution to the total QPI signal, $g_\text{tot}(\mathbf{q}, eV)$. Fig. 2(a) compares calculated QPI maps with and without the contribution from IET induced by spin fluctuations using the T-matrix method (see section S1). An energy independent inelastic tunneling amplitude is assumed in the model. In the five-orbital model, there are three hole bands (labeled as $h_1, h_2, h_3$) and two electron bands (both labeled as $e$) across the Fermi energy. Intra-band and inter-band scattering vectors are highlighted in Fig. 2(a). In contrast to the purely elastic QPI signal, $g_\text{el}(\mathbf{q}, eV)$, the QPI map incorporating the IET effect, $g_\text{tot}(\mathbf{q}, eV)$, shows additional replicas of QPI features close to the original ET features. These replicas are QPI accumulated from $E \leq eV - \Omega$ that is shifted to energy $E$ due to IET. The QPI dispersions of $\mathbf{q}_{h_3-h_3}$ along the Fe-As direction without and with IET are shown in Fig. 2(b) and 2(c), respectively. For $g_{tot}(\mathbf{q}, eV)$, in addition to the normal dispersion of $\mathbf{q}_{h_3-h_3}$, two replica features emerge at both sides of $E_\text{F}$ starting at energy $\pm(\Delta + \Omega_0)$.

**Experimental results**

Fig. 3 shows QPI maps, $g(\mathbf{q}, V)$, measured at different bias voltages. The measurement conditions are given in section S2. The QPI features for $\mathbf{q}_{h_3-h_3}$ and $\mathbf{q}_{h_{1,2}-e}$ scattering are easily identified. On the other hand, $\mathbf{q}_{h_2-h_2}$ lies close to the center in $\mathbf{q}$-space, making it difficult to separate from the low-frequency background. At low bias voltages, the QPI feature corresponding to $\mathbf{q}_{h_3-h_3}$ is a single ring as expected from the QPI simulation without contribution from IET. From 15mV and above, $\mathbf{q}_{h_3-h_3}$ becomes a double ring, with the second ring appearing outside the original ring, as indicated in Fig. 3(d). Comparing to the simulation in Fig. 2, we can attribute the outer ring to the replica feature resulting from IET via bosonic excitations. Here, the most plausible candidate for the bosonic excitation is the spin-fluctuation spectrum [13, 19, 23–26], since no other prominent bosonic mode appears near 15 meV. The energy where replicas emerge also aligns very well with the appearance of the dip-hump structure, indicating a common origin. The replica phenomenon is also observed in the QPI signal associated with $\mathbf{q}_{h_{1,2}-e}$, as highlighted by purple colored arcs in Fig. 3(e), consistent with simulations. In the following, we focus on the analysis of the $\mathbf{q}_{h_3-h_3}$ QPI feature since it is the most well-defined and sharpest pattern in the experimental data.

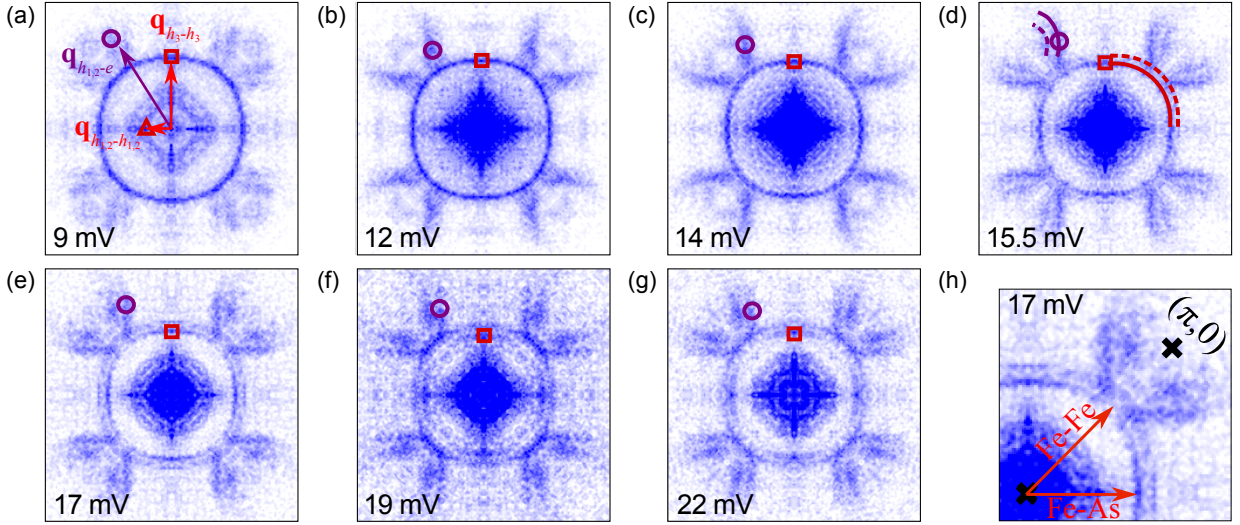

FIG. 3. **Signatures of inelastic tunneling in QPI in momentum space.** (a)-(g) Measured QPI $g(\mathbf{q}, V)$ from $V = 9$ mV to 22 mV at $T = 4.2$ K. QPI features are labeled with triangles for $\mathbf{q}_{h_2-h_2}$, squares for $\mathbf{q}_{h_3-h_3}$, and circles for $\mathbf{q}_{h_{1,2}-e}$ scatterings. In (d), arcs are used to highlight the original (solid) and the replica (dashed) QPI features for $\mathbf{q}_{h_3-h_3}$ and $\mathbf{q}_{h_{1,2}-e}$. (h) Zoomed-in image of $g(\mathbf{q}, 17 \text{ mV})$ with the red arrows indicating the $\mathbf{q}_{h_3-h_3}$ QPI feature along the Fe-Fe and Fe-As directions.

Fig. 4 shows the dispersion of the scattering vector $\mathbf{q}_{h_3-h_3}$ along the Fe-As (Fig. 4a, c) and Fe-Fe (Fig. 4b, d) directions. An integration of the signal over an angle of $\pm \pi/8$ was performed around each direction. Because the QPI due to $\mathbf{q}_{h_3-h_3}$ is almost circular, integration over a small angle is effective for improving the signal-to-noise ratio without distorting the signal. Along the Fe-As direction, $\mathbf{q}_{h_3-h_3}$ can be visualized at both positive and negative bias voltages as indicated by the long dashed line in Fig. 4(a). The replica feature $\mathbf{q}'_{h_3-h_3}$ also appears on both sides of the Fermi energy (0V), as indicated by the short dashed curves. This is consistent with the theoretical simulation shown in Fig. 2(c). Along the Fe-Fe

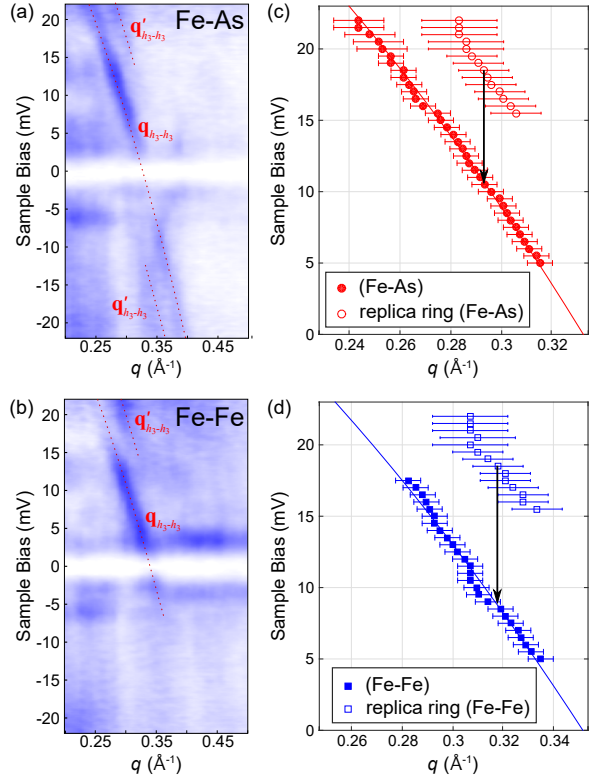

FIG. 4. **Signatures of inelastic tunneling in QPI in dispersions.** (a)-(b) Dependence of $\mathbf{q}_{h_3-h_3}$ QPI along the Fe-As and Fe-Fe directions on the bias voltage $V$. The dashed curves are parabolic fits as guides to the eye. Opening of the superconducting gaps causes fading of the QPI intensities near zero bias. (c, d) Dispersion of the intraband scattering vector $\mathbf{q}_{h_3-h_3}$ extracted in the positive bias range for the Fe-As (c) and Fe-Fe (d) directions. The solid lines are fits of parabolas. The black arrows indicate the direction to shift the IET replicas in order to overlap the original QPI due to elastic tunneling.

direction (Fig. 4(b)), $\mathbf{q}_{h_3-h_3}$ primarily appears in the positive bias side, in agreement with previous experimental observations [27–29]. The intensity of $\mathbf{q}_{h_3-h_3}$ gradually fades away above 15 mV where the replica $\mathbf{q}'_{h_3-h_3}$ becomes more intense. It has been shown that a set-point effect can also lead to additional features [30]. However, features due to the set-point effect in a spectroscopic map do not disperse [30], unlike the replicas we observe here in LiFeAs. Also, the set point effect would not generate a V-shaped tunneling spectrum with *kinks* as seen in Fig. 1 [30]. Therefore, unlike inelastic tunneling, the set point effect cannot explain both the spectral features (Fig. 1) in the tunneling conductance and the replicas in

QPI maps.

This raises the possibility that the *kinks* observed in Ref. [29] are related to the replica features reported here, since they both start at about the same bias voltage $V = 15$mV and have the strongest effect along the Fe-Fe direction [29]. One broader ring instead of the double rings was identified in Ref. [29]. There are a number of possible reasons for this, including different setup conditions or a higher defect density in Ref. [29] resulting in broadening of the QPI so that the elastic QPI and inelastic QPI dispersion are observed as one feature (see section S3).

The dispersion of $\mathbf{q}_{h_3-h_3}$ was extracted from the positive bias side and shown in Fig. 4(c). In both Fe-As and Fe-Fe directions, no obvious abrupt change of the dispersion is noticeable within our resolution, unlike the kink reported by Allan et al. [29]. As indicated by the black arrows, if the replica QPI, $\mathbf{q}'_{h_3-h_3}$, is shifted by 8 mV to 10 mV downward, it aligns well with the dispersion of the original $\mathbf{q}_{h_3-h_3}$ signals. This 8 meV to 10 meV range is a good estimation of the energy of the resonance in the spin-fluctuations and it is consistent with inelastic neutron scattering [24–26] and scanning tunneling spectroscopy results [12, 13, 19, 29].

In the experimental data at higher bias voltages, the intensity of the replica feature $\mathbf{q}'_{h_3-h_3}$ increases, while the intensity of the original feature $\mathbf{q}_{h_3-h_3}$ decreases (see Fig. 3 and Fig. 4). The most dramatic change of the intensity is along the Fe-Fe direction (pointing to $(\pi, \pi)$). While both $\mathbf{q}_{h_3-h_3}$ and $\mathbf{q}'_{h_3-h_3}$ remain along the Fe-As direction above 15mV, the intensity of $\mathbf{q}'_{h_3-h_3}$ becomes dominant along the Fe-Fe direction with increasing bias voltage (see Fig. 3(h)).

While the energy dependence is consistent with simulations (Fig. 2(c)), the angle dependence of the intensity of ET and IET QPI of $\mathbf{q}_{h_3-h_3}$ is absent in the simulation results. This could be due to phase interference effects, particularly due to overlap of the $\mathbf{q}_{h_3-h_3}$ and $\mathbf{q}'_{h_3-h_3}$ with $\mathbf{q}_{h_{1,2}-e}$ in the Fe-Fe direction, however one would expect this to exhibit a sudden suppression/enhancement where the QPI wavevectors overlap rather than the gradual change from the Fe-As to Fe-Fe directions observed (see Fig. 3(e)-(g)). We note that the simulations do not include self-energy effects, which due to correlations can result in an angular dependence of the primary QPI signal. Orbital dependent correlations and quasi-particle weights as proposed, for example, in FeSe[31], may contribute to significant angle-dependence of the elastic QPI features. Self-energy effects would however not be expected to result in replica features.

Alternatively, a momentum dependence of the inelastic coupling to the spin-fluctuations could also give rise to the observed anisotropy of the QPI intensity.

In LiFeAs, the spin-fluctuation spectrum is peaked in the semi-nesting vector between hole and electron bands, aligned with the Fe-Fe direction [13, 24, 25]. In the tunneling model, the momentum dependence of the spin-fluctuation spectrum is integrated out under the approximation of constant tunneling amplitudes [32]. However, the real tunneling matrix will be momentum dependent, given the finite size of the tip [33, 34]. In the ET channel, the momentum dependent tunneling amplitudes can cause the *tunneling cone* effect, favoring the tunneling of states with small in-plane momenta.

Intuitively, with the constraint of a *tunneling cone*, momentum transfer via IET can enhance the probability for tunnelling to states with large in-plane momentum[35]. In this case, IET signals should be stronger along the Fe-Fe direction where the spin-fluctuation spectrum is peaked. This is consistent with angular intensity variation of $\mathbf{q}'_{h_3-h_3}$ seen in experimental data. This naive picture still does not explain the weakened $\mathbf{q}_{h_3-h_3}$ along the Fe-Fe direction at elevated energies. A proper theory that takes the momentum dependence of the tunneling matrix into account is highly desirable for a complete understanding of the angle dependence of $\mathbf{q}_{h_3-h_3}$ and $\mathbf{q}'_{h_3-h_3}$.

**SUMMARY**

We have identified the effect of IET on the QPI dispersion for LiFeAs. This IET effect is expected to exist in many materials with bosonic excitations, in particular where these are electronic in nature, leading to unusual QPI features. In LiFeAs, IET causes replicas of QPI features due to elastic tunneling above 15 meV, echoing the dip-hump structure of the tunneling conductance. The replicas not only give a direct estimation of the resonance energy of spin-fluctuations, but also can potentially be applied to resolve the momentum structure of spin fluctuations, as recently proposed[36]. By including the IET effect, we have a unified understanding of both the spectral features in the local tunneling conductance, as well as their spatial variation as reflected in the QPI dispersion.

ACKNOWLEDGEMENTS

The authors are grateful for helpful conversations with George Sawatzky, Mona Berciu. Research at UBC was supported by the Natural Sciences and Engineering Research Council, the Canadian Institute for Advanced Research, and the Canadian Foundation for Innovation. CDAM and PW acknowledge support through the LIFTS scheme of the University of St Andrews, and PW from the Leverhulme Trust through RPG-2022-315.

———

* deceased 11 May 2025

† wahl@st-andrews.ac.uk

## S1. SUPPLEMENTARY: THEORY

A five-orbital tight-binding model $H_0$ for LiFeAs from Ref. [37] is used for QPI simulation. There are five bands across the Fermi energy ($E_F$) as shown in Fig. 5(a). These are three hole bands $h_1, h_2$ and $h_3$ and two electron bands. In the model, two electron bands are identical except for their locations in the Brillouin zone, $(\pm\pi, 0)$ vs. $(0, \pm\pi)$. Therefore one common label $e$ is used for both electron bands.

The total Hamiltonian including defect scattering is given by

$$H = H_0 + H_{\text{imp}}, \tag{1}$$

where $H_0$ is the bare Hamiltonian including the superconducting BCS term,

$$H_0 = \begin{bmatrix} \epsilon(\mathbf{k}) & \Delta(\mathbf{k}) \\ \Delta^\dagger(\mathbf{k}) & -\epsilon^{\mathrm{T}}(-\mathbf{k}) \end{bmatrix}, \quad (2)$$

with $\epsilon(\mathbf{k})$ representing the bandstructure, and $\Delta(\mathbf{k}) = \Delta_0 \cos(k_x)\cos(k_y)$ being the $s_\pm$ order parameter with $\Delta_0 = 0.014$ eV. The scattering at a point-like defect is introduced through $H_{\mathrm{imp}}$, given by

$$H_{\mathrm{imp}} = V_0 \sum_{\mu\sigma} c^\dagger_{0\mu\sigma} c_{0\mu\sigma}, \quad (3)$$

which describes the scattering for orbital $\mu$ and spin $\sigma$ at site $\mathbf{r} = (0,0)$ with potential $V_0$. Only intra-orbital scattering is considered in the calculation because it leads to the dominant components in the scattering matrix [37]. For simplicity, equal scattering potential across all orbitals is assumed in this study. The $T$-matrix is calculated from the equation $T(\omega) = [I - \frac{V_0}{N}\sum_\mathbf{k} G_0(\mathbf{k},\omega)]^{-1} \frac{V_0}{N}$.

The Green's function in real space can be obtained from the $T$-matrix

$$G(\mathbf{r},\mathbf{r}',\omega) = G_0(\mathbf{r},\mathbf{r}',\omega) + NG_0(\mathbf{r},\mathbf{0},\omega)T(\omega)G_0(\mathbf{0},\mathbf{r}',\omega), \quad (4)$$

where $G_0(\mathbf{r},\mathbf{r}',\omega)$ is the Fourier transformation of $G_0(\mathbf{k},\omega)$. The LDOS is given by

$$\begin{aligned} \rho(\mathbf{r},\omega) &= -\frac{1}{\pi}\mathrm{Im}G(\mathbf{r},\mathbf{r},\omega) \\ &= -\frac{1}{\pi}\mathrm{Im}G_0(\mathbf{r},\mathbf{r},\omega) - \frac{1}{\pi}\mathrm{Im}NG_0(\mathbf{r},\mathbf{0},\omega)T(\omega)G_0(\mathbf{0},\mathbf{r},\omega) \\ &= \rho_0(\mathbf{r},\omega) + \delta\rho(\mathbf{r},\omega). \end{aligned} \quad (5)$$

Finally, the QPI intensity map of elastic tunneling is given by

$$g_{\mathrm{el}}(\mathbf{q},V) \propto \delta\rho(\mathbf{q},\omega=eV) = \frac{1}{N}\sum_\mathbf{r} e^{-i\mathbf{q}\cdot\mathbf{r}}\delta\rho(\mathbf{r},\omega). \quad (6)$$

We utilized Eq. 2 from Ref [32] to derive the IET contribution to the tunneling conductance $g^{\mathrm{inelastic}}(\omega)$. In the limit where $k_\mathrm{B}T \ll eV$, the derivative of the Fermi-Dirac distribution function approximates a delta function, and the Bose-Einstein distribution function is only significant at 0 meV. Given that the spin-fluctuation strength is zero at 0 meV, Eq. 2 from Ref [32] simplifies to:

$$\begin{aligned} g^{\mathrm{inelastic}}(\omega,\mathbf{r}) = C\Big[&\int_{-\infty}^{\infty} n_\mathrm{F}(\omega')\mathrm{Im}\chi(\omega'-\omega)\rho_s(\omega',\mathbf{r})\mathrm{d}\omega' \quad (7)\\ &+ \int_{-\infty}^{\infty}(1-n_\mathrm{F}(\omega'))\mathrm{Im}\chi(\omega-\omega')\rho_s(\omega',\mathbf{r})\mathrm{d}\omega'\Big], \quad (8) \end{aligned}$$

where $\text{Im}\chi(\omega)$ is the imaginary part of the spin susceptibility and $\rho_s(\omega)$ is the sample DOS. The constant prefactor $C$ is determined by the quantum yield of the excitation of the mode during the tunneling process.

We use $\text{Im}\chi(\omega)$ from Ref 13, re-scaled according to the gap sizes to simulate the effect of inelastic tunneling (see the insert of Fig. 1). The prefactor was determined to match experimental V-shaped spectrum. We then calculate the QPI from equation 6 in real space and use equation 8 to add the inelastic contribution to simulate the QPI intensity map including the effect of inelastic tunneling before taking the Fourier transformation to obtain $g_{\text{tot}}(\mathbf{q}, V)$.

Fig. 5 shows the Fermi surface derived from the model and QPI above the superconducting gaps. QPI scattering vectors are highlighted using red and purple arrows. Here the images are rotated by $90°$ to match the orientation of the experimental data (Fig. 3 in the main text). $\mathbf{q}_{h_1-h_1}$ is the tiny ring merged in a blurred pie-like signal. It is also hard to discern $\mathbf{q}_{h_1-h_1}$ from experimental data. Thus, $\mathbf{q}_{h_1-h_1}$ is not indicated in the graph.

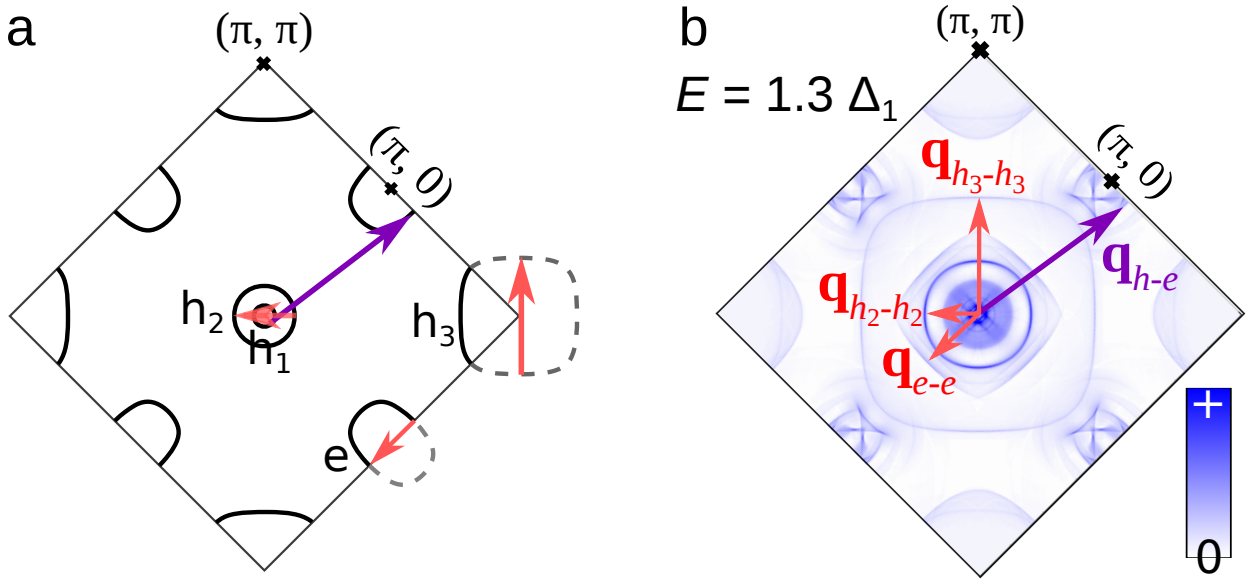

FIG. 5. (a) Fermi surface of LiFeAs derived from the five-orbital model. Red and purple arrows indicate possible QPI features, corresponding to intra-band and inter-band scattering processes. (b) Simulated QPI above the superconducting gap ($E = 1.3\Delta_1$) with prominent features labeled $\mathbf{q}_{h_2-h_2}$, $\mathbf{q}_{h_3-h_3}$, $\mathbf{q}_{e-e}$, and $\mathbf{q}_{h-e}$.

## S2. SUPPLEMENTARY: EXPERIMENTAL DETAILS

High purity single crystals of LiFeAs ($T_c = 17.2$ K, $\Delta T_c = 1$ K) were grown by a self-flux technique [19]. For the STM/STS measurements, one LiFeAs single crystal sample was cleaved in-situ at cryogenic temperature and inserted into an STM.

The experimental tunneling conductance spectrum shown in Fig. 1(b) was measured in a home-built STM operating at temperatures down to 1.5K [38]. QPI data was acquired in a beetle-type Createc STM at the base temperature 4.2 K. Data was taken with a grid size $400 \times 400$ on a $26 \times 26$ nm$^2$ area, as shown in Fig. 6(a). The set-up conditions are $I_{set} = 250$ pA and $V_{set} = 25$ mV. At each pixel, an $I$-$V$ spectrum with 512 data points was acquired. The tunneling conductance data were obtained by numerical differentiation of $I$-$V$ spectra. Fig. 6(c) shows a typical $g(V)$ spectrum derived from the grid at a defect-free location. Due to the thermal broadening effect at 4.2 K, only the large superconducting gap with $\Delta = 6$ meV is discernible. In the tunneling conductance maps, for example, $g(\mathbf{r}, V = 8\,\text{mV})$ shown in Fig. 6(d), spatial oscillations due to QPI near defects are clearly visible.

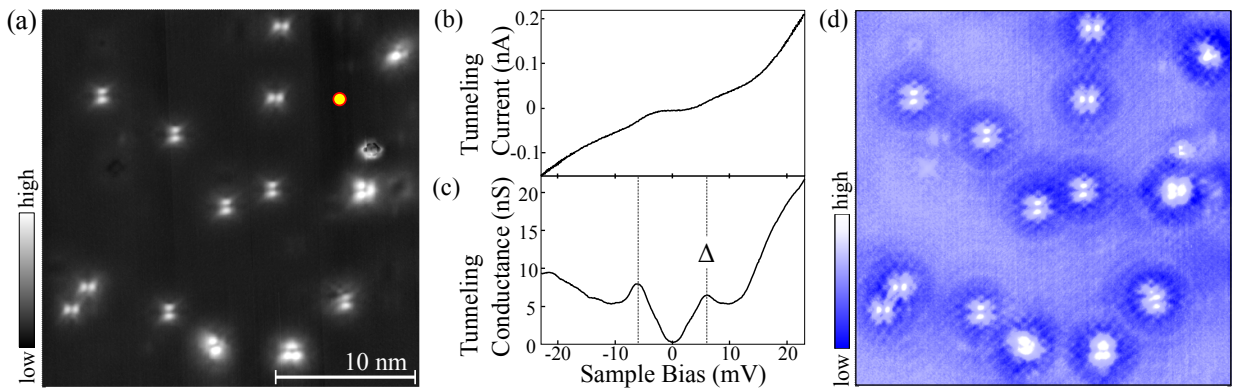

FIG. 6. (a) Topography of the area for the QPI measurement. (b) An $I$-$V$ spectrum at the defect-free location marked by a yellow dot in (a). (c) $g(V)$ obtained by numerical differentiation of the $I$-$V$ spectrum in (b). (d) The tunneling conductance map $g(\mathbf{r}, V = 8\,\text{mV})$.

## S3. SUPPLEMENTARY: SIGNAL TO NOISE RATIO

In a QPI measurement, several factors can influence the Signal-to-Noise Ratio (SNR), including material-intrinsic properties and electronic noise. This discussion focuses on two critical experimental setting parameters: the field of view (measurement area) $A$ and the number of pixels $M$. For the purpose of this analysis, we assume a low defect density within the QPI measurement area. This assumption allows us to neglect the multi-scattering effect. It is noteworthy that defects in LiFeAs exhibit strong non-QPI peak/trough signals at their centers, which contribute to an overall background in $\mathbf{q}$-space, thereby obscuring the true QPI signals [28]. Therefore, maintaining a low defect density is crucial for reducing this noise background.

QPI modulations in a tunneling conductance map with multiple defects can be expressed as

$$g(\mathbf{r},\omega) = \sum_{i}^{N} g_S(\mathbf{r} - \mathbf{R}_i, \omega), \tag{9}$$

where $\mathbf{R}_i$ is the location of the $i$-th defect and $g_S(\mathbf{r}, \omega)$ is the QPI oscillations for a single defect at the origin. The QPI amplitude in $\mathbf{q}$-space is

$$\begin{aligned}
|g(\mathbf{q},\omega)| &= |\int d\mathbf{r}\, e^{-i\mathbf{q}\mathbf{r}} g(\mathbf{r},\omega)| \\
&= |\int d\mathbf{r}\, e^{-i\mathbf{q}\mathbf{r}} \sum_{i}^{N} g_S(\mathbf{r} - \mathbf{R}_i, \omega)| \\
&= |\sum_{i}^{N} e^{i\mathbf{q}\mathbf{R}_i} \int d\mathbf{r}\, e^{-i\mathbf{q}\mathbf{r}} g_S(\mathbf{r},\omega)| \\
&= |\sum_{i}^{N} e^{i\mathbf{q}\mathbf{R}_i}||g_S(\mathbf{q},\omega)|,
\end{aligned} \tag{10}$$

where $g_S(\mathbf{q},\omega)$ is the QPI signal of a single defect in $\mathbf{q}$-space. From Eq. 10, the strength of the QPI signal in $\mathbf{q}$-space measured across an area with $N$ defects is not $N$ times the QPI signal from a single defect. The factor $|\sum_{i}^{N} e^{i\mathbf{q}\mathbf{R}_i}|$ adds further modulation to the signal from a single defect. As a result, the random distribution of defects gives on average,

$$|g(\mathbf{q},\omega)| \propto |\sum_{i}^{N} e^{i\mathbf{q}\mathbf{R}_i}| \propto \sqrt{N} \propto \sqrt{nA}, \tag{11}$$

where $n$ is the defect density of the material and $A$ is the measurement area. Thus the total signal gain in $\mathbf{q}$-space from multiple defects is proportional $\sqrt{nA}$.

In Eq. 10, the second part on the right hand side is the QPI signal of a single defect, $g_S(\mathbf{q},\omega)$. In a typical measurement, data were acquired on an evenly distributed discrete grid with a fixed number of pixels, $M$. $g_S(\mathbf{q},\omega)$ is given by the discrete Fourier transformation

$$g_S(\mathbf{q},\omega) = \sum_i^M e^{-i\mathbf{q}\mathbf{r}_i} g_S(\mathbf{r}_i,\omega). \tag{12}$$

Given an area $A$ and $M$ pixels, the pixel resolution in $\mathbf{r}$-space is $\frac{M}{A}$.

If the pixel resolution is doubled in the same area, for example, taking measurement in the middle of each unit grid box as well,

$$g_S(\mathbf{q},\omega) = \sum_i^M e^{-i\mathbf{q}\mathbf{r}_i} g_S(\mathbf{r}_i,\omega) + \sum_i^M e^{-i\mathbf{q}(\mathbf{r}_i+\delta\mathbf{r})} g_S(\mathbf{r}_i+\delta\mathbf{r},\omega). \tag{13}$$

where $\delta\mathbf{r}$ is the shift distance between the original grid and the second grid. $\delta\mathbf{r}$ is typically much smaller than the inverse of scattering vectors $\mathbf{q}$ in the first Brillouin zone, giving $e^{-i\mathbf{q}\delta\mathbf{r}} \approx 1$ and $g_S(\mathbf{r}_i+\delta\mathbf{r},\omega) \approx g_S(\mathbf{r}_i,\omega)$.

Therefore,

$$\begin{aligned} g_S(\mathbf{q},\omega) &= \sum_i^M e^{-i\mathbf{q}\mathbf{r}_i} g_S(\mathbf{r}_i,\omega) + e^{-i\mathbf{q}\delta\mathbf{r}} \sum_i^M e^{-i\mathbf{q}\mathbf{r}_i} g_S(\mathbf{r}_i+\delta\mathbf{r},\omega) \\ &\approx \sum_i^M e^{-i\mathbf{q}\mathbf{r}_i} g_S(\mathbf{r}_i,\omega) + \sum_i^M e^{-i\mathbf{q}\mathbf{r}_i} g_S(\mathbf{r}_i+\delta\mathbf{r},\omega) \\ &\approx \sum_i^M e^{-i\mathbf{q}\mathbf{r}_i} g_S(\mathbf{r}_i,\omega) + \sum_i^M e^{-i\mathbf{q}\mathbf{r}_i} g_S(\mathbf{r}_i,\omega) \\ &= 2\sum_i^M e^{-i\mathbf{q}\mathbf{r}_i} g_S(\mathbf{r}_i,\omega). \end{aligned} \tag{14}$$

Eq. 14 shows that the amplitude of $g_S(\mathbf{q},\omega)$ is approximately doubled when the pixel resolution is doubled. Thus, we have

$$|g_S(\mathbf{q},\omega)| \propto \frac{M}{A}. \tag{15}$$

For a fixed random noise spectrum, the noise strength in $\mathbf{q}$-space is proportional to $\sqrt{M}$. The SNR is related to the field of view and the number of pixels as

$$\mathrm{SNR}(|g(\mathbf{q},\omega)|) \propto \frac{\sqrt{nA} \times \frac{M}{A}}{\sqrt{M}} = \sqrt{\frac{nM}{A}}. \tag{16}$$

Eq. 16 shows that given a fixed impurity density $n$, increasing pixel resolution $\frac{M}{A}$ improves SNR.

In a comparative analysis of the $\mathbf{r}$-space QPI maps between Fig. 6(c) and Figure-S5 in Ref [29], it is clear that our data offers a superior pixel resolution. Specifically, we have achieved a resolution of 237 pixels/nm$^2$ over an area of 26×26 nm$^2$. This enhanced resolution is a primary factor in the improved SNR, facilitating the resolution of the IET scattering vectors with greater precision.

In addition, the region under investigation exhibits an optimal impurity density, which significantly contributes to the enhancement of the SNR. Importantly, the impurity level is not excessively high, thereby facilitating the clear observation of Friedel oscillations between the impurities, as shown in Fig. 6(c). Intrinsic impurities in LiFeAs are characterized by prominent peak/trough signals at their centers, which are distinct from Friedel oscillations. These signals introduce a considerable noise background in **q**-space following Fourier transformation. To mitigate the noise contribution from defect centers, we employed a Gaussian masking method [28]. However, it should be noted that an excessively high impurity density complicates the separation of Friedel oscillations from impurity center signals, potentially compromising the SNR in **q**-space.

## S4.  SUPPLEMENTARY: QPI FOR ALL ENERGIES

The QPI maps between -22 meV and 22 meV are shown in *QPIallEnergy.mp4*.